\documentclass[9pt,twocolumn,twoside,lineno]{pnas-new}

\articletype{CLASSIFICATION}%%%% Article topic/classification

\templatetype{pnasresearcharticle}

\usepackage[utf8]{inputenc}
\usepackage[T1]{fontenc}
\usepackage{textcomp}

\usepackage{lineno}

\usepackage{amsmath,amssymb,amsfonts}
\usepackage{bm}
\usepackage{dcolumn}
\usepackage{etoolbox}
\usepackage{float}
\usepackage{graphicx}
\usepackage{xcolor}
\usepackage{url}

\usepackage{hyperref}

\setboolean{displaywatermark}{false}

\begin{document}

% \title{Evolving Neural Networks Reveal Emergent Collective Behavior from Minimal Agent Interactions}
\title{Evolved Collectives Combine Complex Internal Representations with Simple Outputs}

\author[a,1]{G. ~S. ~Y. ~Giardini}
\author[a]{J. ~F. ~Hardy II}
\author[b,c]{C. ~R. ~daCunha}

\affil[a]{School of Informatics, Computing, and Cyber-Systems, Northern Arizona University, Flagstaff, AZ 86011 USA.}%
\affil[b]{Helen and John C. Hartmann Department of Electrical and Computer Engineering, New Jersey Institute of Technology, 323 Martin Luther King Jr. Blvd, Newark, NJ 07102, USA}
\affil[c]{School of Applied Engineering and Technology, New Jersey Institute of Technology, 323 Martin Luther King Jr. Blvd, Newark, NJ 07102, USA}

\leadauthor{Giardini} 

\significancestatement{When animal groups or robot swarms coordinate, each individual makes decisions using only limited local information. We asked how the neural controllers behind this coordination are organized internally. As  collective order increases, the controllers do not become uniformly more complex. Their internal representations grow richer, while their output decisions become simpler and more linear. Even random, untrained controllers with more linear outputs can produce more ordered groups, suggesting that collective environments favor simplicity at the decision stage. Evolution then sharpens this pattern. The result is a division of structure inside each controller: richer representation where collective information must be organized, simpler mapping where action is produced.
}

%Please include corresponding author, author contribution and author declaration information
% \authorcontributions{Please provide details of author contributions here.}
% \authordeclaration{Please declare any competing interests here.}
\correspondingauthor{\textsuperscript{1} G.~S.~Y.~Giardini (gsg95@nau.edu).}


\begin{abstract}
Collective intelligence emerges from local interactions among agents with limited information, yet how internal controller organization relates to emergent collective order remains unclear. Here, we study evolved swarms with shallow neural controllers under explicit sensory and actuation constraints and compare collective order with hidden-layer complexity and output nonlinearity across 3024 conditions. Under these constraints, the most ordered regimes exhibit two simultaneous and seemingly contrasting effects: hidden-layer complexity increases, while the effective output mapping becomes more linear. The diversity of recurrent collective behaviors varies nonmonotonically across the control parameters, with pattern richness shaped by parameter-specific tradeoffs rather than a single generic constraint optimum. Unevolved controls show that output linearization persists without adaptation, whereas the hidden-complexity relation depends on optimization. These two effects are respectively consistent with the law of requisite complexity and ecological rationality, suggesting that adaptive collective intelligence can arise through a partitioned controller organization in which representational complexity and action-level linearization coexist within the same system.
\end{abstract}

% \dates{This manuscript was compiled on \today}
\dates{\textcolor{red}{Preprint manuscript. This version has not been peer reviewed or accepted for publication.} \\
This manuscript was compiled on \today.}
% \doi{\url{www.pnas.org/cgi/doi/10.1073/pnas.XXXXXXXXXX}}
\doi{\textcolor{red}{Preprint manuscript. \\ This version has not been peer reviewed or accepted for publication.}}

\maketitle
\thispagestyle{firststyle}
\ifthenelse{\boolean{shortarticle}}{\ifthenelse{\boolean{singlecolumn}}{\abscontentformatted}{\abscontent}}{}

\Firstpage

\label{sec:Introduction}
Collective systems from animals to robots generate large-scale organization from local interactions, even though they comprise agents that possess limited information about their surroundings \cite{Ballerini2008,Giardina2008,Gautrais2012,Kumar2021}. Across these systems, coherent spatial patterns emerge from repeated local decisions and persist over large spatial and temporal scales, suggesting that collective intelligence may be governed by organizational principles linking local information processing to macroscopic behavior \cite{Bialek2012,Charlesworth2019,Suran2020}.

A central missing piece is the internal computation of the controllers that generate collective order. Prior work has characterized emergent patterns, order parameters, and phase structure in distributed systems \cite{Reynolds1987,Vicsek1995,OlfatiSaber2004,Costanzo2018}, while separate lines have studied neural and evolutionary controllers in terms of adaptation, optimization, and performance \cite{Yao1999,Nolfi2000,Kwasnicka2011,Witkowski2016,Ramos2019}. What remains less clear is how emergent spatial organization relates to the internal representational structure and output mappings of the controllers themselves. If collective behavior depends on adaptive decision-making, then understanding how those computations are organized is part of understanding collective intelligence \cite{Suran2020,Holland1998,Yao1999}.

Evolutionary swarms provide a useful setting for this question because they couple adaptive local decision-making to macroscopic collective behavior within a controlled framework. Each agent acts from a restricted sensory neighborhood, transforms that information through a neural controller, and contributes to a global pattern that no individual observes in full, making the system suitable for studying how limited information and embodiment shape adaptive control \cite{Trianni2008,Kwasnicka2011,Brambilla2013,giardini2025feardriven}.

Here, we study an evolutionary swarm framework in which shallow neural controllers \cite{Cybenko1989} evolve under explicit sensory and actuation constraints \cite{Costanzo2018}. We organize the resulting behaviors with an interpretable geometric \Parasplit classifier and characterize the controllers through complementary descriptors of hidden-layer representation, output mapping, and sensory-environment complexity. This design allows us to relate collective order, controller organization, and the range of collective regimes supported under different constraint conditions.

We show that informational constraints shape both the collective organization and the internal organization of the evolving controllers. Regime richness varies nonmonotonically across the control parameters, with moderate neighborhood size, low but nonminimal noise, and more edge-weighted values of field of view and turning capacity supporting the largest pattern counts. Ordered regimes are associated with higher hidden-layer complexity and more linear output mappings. Hidden-layer complexity covaries with the structured complexity of the local sensory environment, consistent with the law of requisite complexity \cite{Ashby1956,Boisot2011}, whereas the output stage satisfies the central prediction of Gigerenzer's ecological rationality framework: under informational constraints, more linear decision structure is associated with stronger collective order \cite{Gigerenzer1999}. Together, these findings link emergent spatial organization to internal controller architecture and suggest that distributed intelligence may arise through a division between internally differentiated representation and action-level simplification.

\section*{Model}
\label{sec:Model}

\subsection*{System Overview}

We consider a decentralized multi-agent system in which collectivity emerges from local interactions among agents with limited information and bounded action. Agents move in a two-dimensional domain with fixed speed and update their heading direction from locally available sensory inputs alone. Each agent is controlled by a shallow neural controller that maps relative positional information from nearby agents to changes in orientation, allowing collective organization and controller structure to coemerge under adaptation.

Interaction and control are constrained through limited sensing, bounded turning rate, stochastic perturbations, collision avoidance, and a fixed number of interacting neighbors. These quantities define the informational and actuation limits of the system and serve as the primary experimental variables of the study. By varying them systematically, we examine how constraint shapes both the range of recurrent collective regimes and the controller organizations associated with ordered behavior. The full system is formalized below.

\subsection*{Agent State and Environment Model}

The system consists of $N$ agents with positions
{
\(
\mathbf{X} = \{\mathbf{x}_{1},\mathbf{x}_{2}, \dots, \mathbf{x}_{N}\},
\)
}
where each agent position $\mathbf{x}_{i}(t)$ lies in a finite two-dimensional domain $\mathcal{L} \subset \mathbb{R}^{2}$ of size $L \times L$ within periodic boundaries. Each agent also possesses an orientation $\theta_{i}(t)$ that determines its direction of motion. 

These agents move at a fixed speed $v_{0}$ and occupy a finite volume enforced through a collision radius $R_{\text{col}}$, which prevents multiple agents from occupying the same spatial location. 

\subsection*{Local Neighborhood and Sensing}

Each agent interacts with a ranked metric neighborhood
\(
\mathcal{N}_{i} = \{\mathbf{y}_{1}, \mathbf{y}_{2}, \dots, \mathbf{y}_{\kappa}\},
\)
consisting of the $\kappa$ nearest neighboring agents according to a distance metric $d(\cdot, \cdot)$, such that
\(
d(\mathbf{x}_{i},\mathbf{y}_{1}) < d(\mathbf{x}_{i},\mathbf{y}_{2}) < \dots < d(\mathbf{x}_{i},\mathbf{y}_{\kappa})
\)
defines the neighboring rank. 

\begin{figure}%[tbhp]
    \centering
    \includegraphics[width=0.9\linewidth]{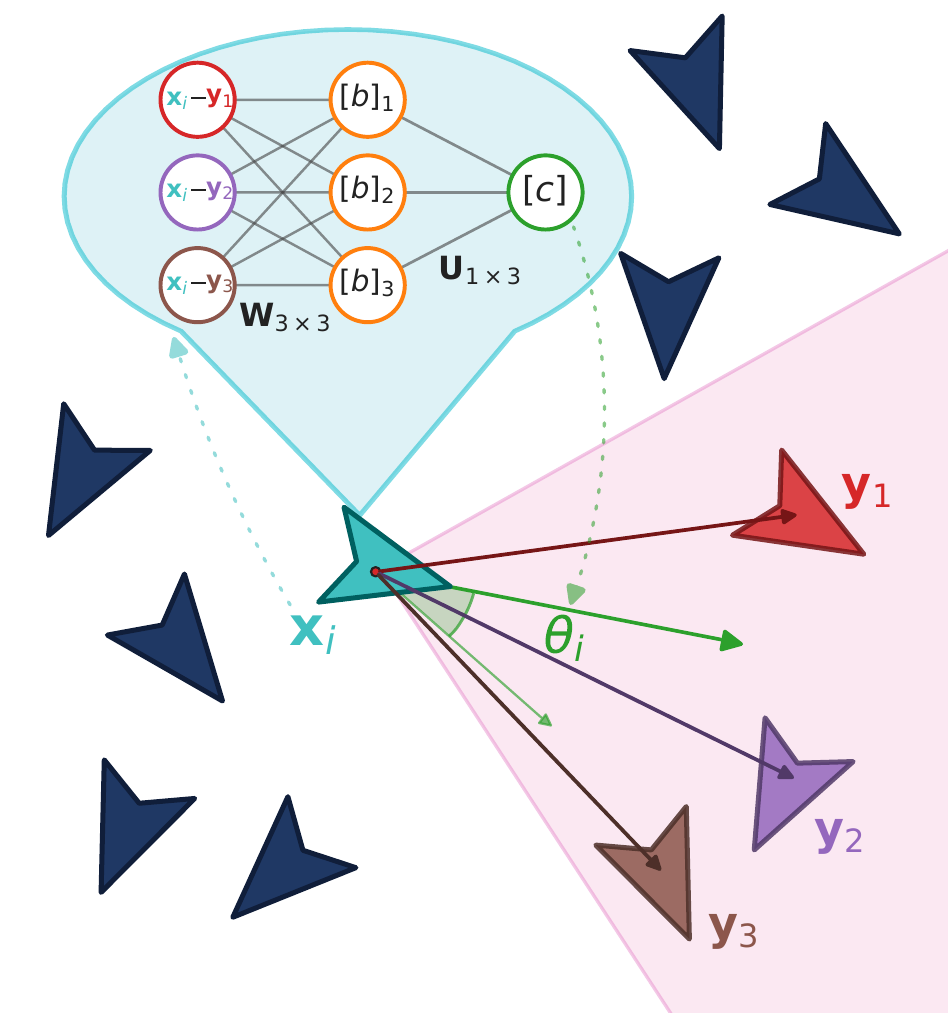}
    \caption{Simplified diagrammatic representation of interacting agents with 3 nearest neighbors. In the diagram, the number of nearest neighbors has been defined as $\kappa=3$ for simplicity. The variables $\mathbf{x}_{i}$ and $\theta_{i}$ are respectively agent $i$'s position and orientation, the pink sector of a circle represents its field of view. The variables $\mathbf{y}_{j}$ are the positions of neighboring agents inside agent $i$'s field of view, and the blue region denotes its neural network, with inputs $\mathbf{v}_i = [\mathbf{x}_{i}-\mathbf{y}_{j}]_{j=1,\hdots,\kappa}$, a weight matrix $\mathbf{W}$ with bias vector $\mathbf{b}$ whose result passes through an activation function $\tanh$ and is fed into an output layer with a weight vector $\mathbf{u}$ and activation value $c$.}
    \label{fig:sec2}
\end{figure}

Only neighbors within an agent's field of view $\phi \in [0, 2\pi]$ are considered as inputs. This restricts sensing to a local and directional subset of the environment. A schematic illustration of the interaction geometry and controller inputs is shown in Fig. \ref{fig:sec2}.

\subsection*{Neural Controller}

Each agent is controlled by a shallow feedforward neural network that maps local relative positional information to changes in orientation. The input to the network for an agent at position $\mathbf{x}_{i}$ is a vector of relative displacement
\begin{equation}
\nonumber \mathbf{v}_i = [\mathbf{x}_{i} - \mathbf{y}_{j}]_{j=1,\hdots,\kappa}, 
\end{equation}
where $\mathbf{y}_{j}$ belongs to agent $i$'s neighborhood $\mathcal{N}_{i}$. 

The network consists of a single hidden layer with eight neurons, 
\begin{equation}
\nonumber \mathbf{h} = \tanh(\mathbf{W}\mathbf{v} + \mathbf{b}),
\end{equation}
followed by an output layer 
\begin{equation}
\nonumber Y = \tanh(\mathbf{u}\cdot \mathbf{h} + c)
\end{equation}
where $\mathbf{W}$, $\mathbf{u}$, $\mathbf{b}$, and $c$ are trainable parameters. The scalar output $Y$ determines the agent's desired change in heading $\theta$. 

\subsection*{Motion Update and Embodiment Constraints}

The neural controller produces a desired angular increment
\begin{equation}
\nonumber \Delta\theta^{\ast} = (1-\lambda)\,Y\pi + \lambda\,\xi_t,
\end{equation}
where $\lambda\in[0,1]$ controls the noise amplitude and $\xi_t$ is a Wiener process \cite{Wiener1923,Genthon2020,daCunha2022}.

The agent's orientation is then updated according to
\begin{equation}
    \theta(t+\Delta_{t}) =
    \theta(t) +
    \begin{cases}
    \Delta\theta^{\ast}, & \text{if } |\Delta\theta^{\ast}| \le \omega_{\text{max}} \Delta_{t}, \\
    \operatorname{sgn}(\Delta\theta^{\ast})\,\omega_{\text{max}} \Delta_{t}, & \text{otherwise},
    \end{cases}
\end{equation}
where $\omega_{\text{max}}$ is the maximum angular velocity \cite{Costanzo2018}.

The updated orientation determines the agent's position update
\begin{equation}
    \nonumber \mathbf{x}(t+\Delta_t)=\begin{cases}\mathbf{x}(t)+\begin{bmatrix}\cos(\theta)\\ 
    \sin(\theta)\end{bmatrix}v_0\Delta_t & \text{, if no collision occurs}\\
    \mathbf{x}(t) & \text{, otherwise.}\end{cases}
\end{equation}
These constraints enforce bounded actuation, stochasticity, and volume exclusion.

\subsection*{Evolutionary Dynamics and Fitness}

The neural controller parameters are optimized online using an evolutionary algorithm. Every $T=300$ simulation steps, agents are selected via fitness-proportionate selection \cite{Lipowski2012}. Selected parameter vectors are recombined via uniform crossover \cite{daCunha2024} and subject to mutation, while offspring replace the least-fit agents through an elitist replacement strategy \cite{Holland1998,Yao1999,FLOREANO2023,daCunha2024}. This process continues until $20\%$ of the population has reproduced, after which the evolutionary cycle restarts.

Fitness itself is defined as the time-averaged inverse distance between an agent and its $\kappa$ nearest neighbors,
\begin{eqnarray}
    \varphi_{i}(t+\Delta_t) = \frac{1}{t+\Delta_t} \left[t \, \varphi_{i}(t) + \frac{\Delta_t}{k} \sum_{j\in \mathcal{N}_{i}} |
\mathbf{x}_i-\mathbf{x}_j|^{-1} \right].
    \label{eq:Sec_2_fitness}
\end{eqnarray}
To maintain controller diversity and reduce premature convergence, fitness is further adjusted using a sharing correction following
\cite{FLOREANO2023}. The full sharing formulation and parameterization are provided in SI Appendix.

All neural networks are initialized randomly, and training proceeds until a steady state is reached at approximately $1.5 \times 10^{5}$ simulation steps. The evolved controllers and their resulting collective configurations are then used as the basis for the multi-level analysis described in the following section.

\section*{Results}
\label{sec:Results}

\subsection*{Sensory Constraints Shape the Collective Regime Space}

\begin{figure*}[!t]
      \centering
      \includegraphics[width=\linewidth]{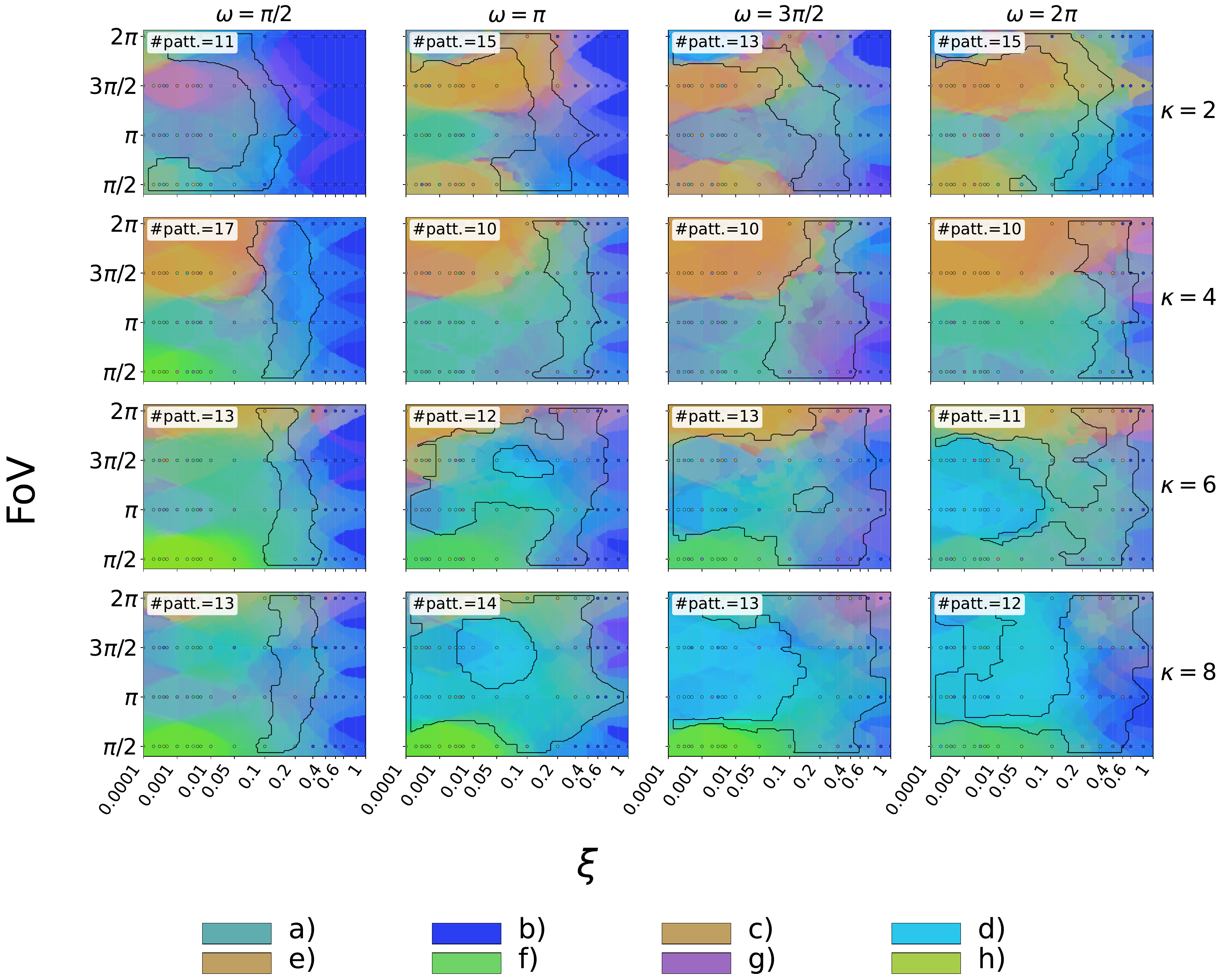}
      \caption{Classifier-defined phase map over the full 3024-run parameter sweep. Colored points denote simulated parameter combinations labeled by the dominant recurrent regime. The surrounding phase field is obtained by interpolation. Colors encode phase identity rather than numerical magnitude. The reference labels correspond to the following representative regime families: (a) single elongated unpolarized flowing state, (b) uniform unpolarized state, (c) multiple elongated mixed or unpolarized flowing states, (d) single round unpolarized flowing state, (e) multiple elongated unpolarized flowing states, (f) single elongated nematic-ordered flowing state, (g) single elongated unpolarized stationary state, and (h) single round unpolarized stationary state. Continuous color gradients indicate smooth transitions; black lines denote interpolation boundaries between distinct regimes. $\#\mathrm{patt.}$ denotes the number of distinct patterns per subplot. Columns: $\omega=\{\pi/2,\pi,3\pi/2,2\pi\}$; rows: $\kappa=\{2,4,6,8\}$.}
      \label{fig:phase_space_classification}
  \end{figure*}

We first organize the 3024-simulation ensemble into recurrent collective regimes using an interpretable geometric classifier that assigns compositional labels based on density structure, orientational order, motion state, shape, and propagation direction, described in the Materials and Methods section. The resulting phase map (Fig.~\ref{fig:phase_space_classification}) reveals how the evolved collective behaviors are distributed across the explored parameter space. Each subplot corresponds to a different combination of maximum turning angle $\omega$ and neighborhood size $\kappa$, with colored points marking the dominant regime at each simulated parameter combination. Continuous color gradients between neighboring points indicate smooth regime transitions.

A key feature visible in Fig.~\ref{fig:phase_space_classification} is that regime richness varies nonmonotonically across the control parameters. The pattern counts ($\#$patt.) are largest at moderate neighborhood size, at low but nonminimal noise, and at more edge-weighted values of field of view and turning capacity. Along the neighborhood axis, the maximum occurs at $\kappa=6$, which coincides with the topological neighbor counts often reported in bird flocks. This correspondence is suggestive and may be relevant to the maintenance of flexible collective repertoires. A fuller analysis of pattern counts across the sweep is provided in SI Appendix.

\begin{figure*}[!t]
  \centering
      \includegraphics[width=0.9\linewidth]{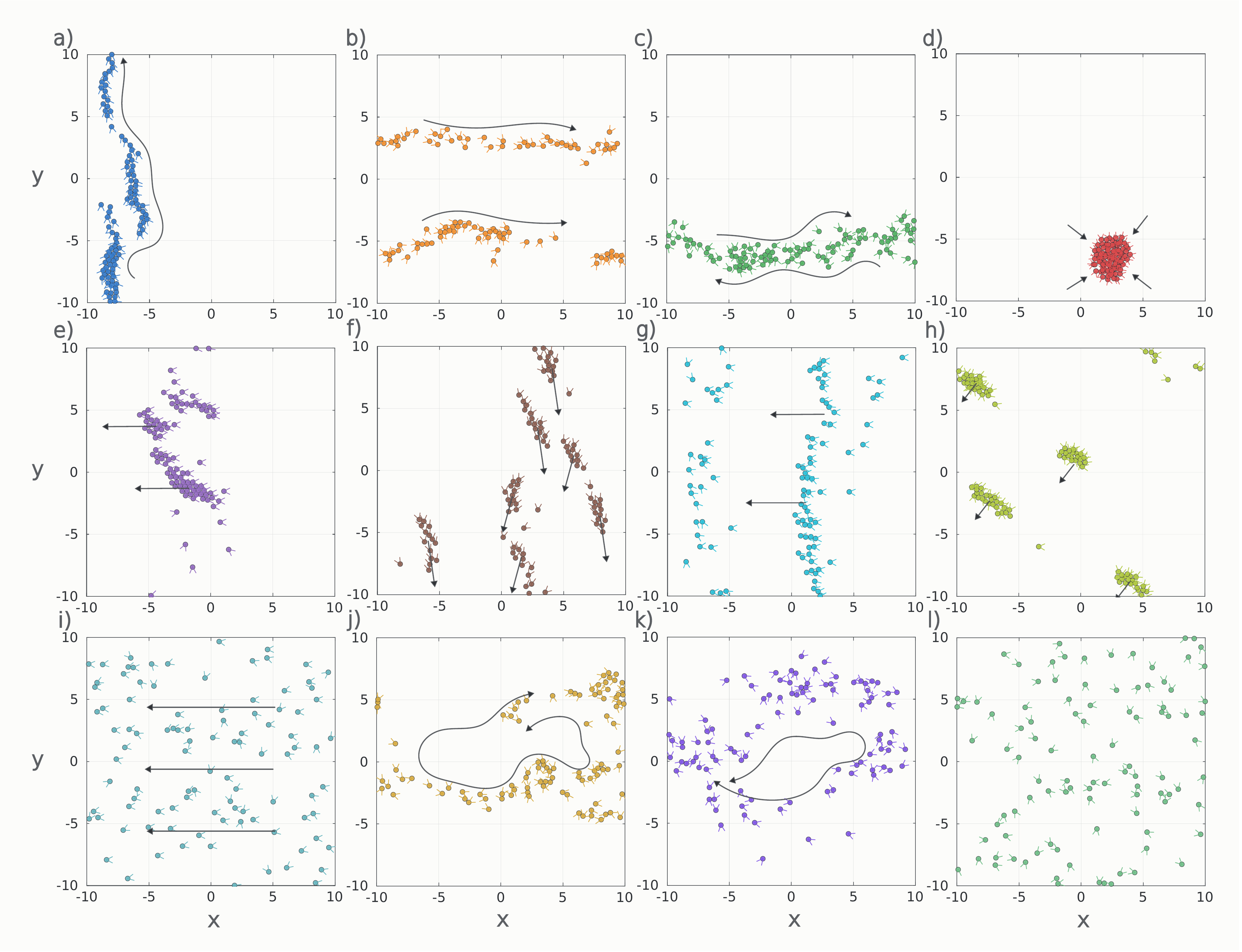}
      \caption{Representative classifier-defined collective regimes: (a) single elongated polarized flow along the major axis, (b) multiple elongated polarized flows along the major axis, (c) single elongated unpolarized stationary state, (d) single round unpolarized stationary state, (e) multiple elongated polarized flows with ambiguous principal-axis propagation, (f) multiple elongated polarized flows along the major axis, (g) single elongated polarized flow along the minor axis, (h) multiple elongated polarized flows along the minor axis, (i) uniform polarized state, (j) single round unpolarized flowing state, (k) single round unpolarized stationary state, and (l) uniform unpolarized state. Black arrows indicate aggregate direction of motion.}
      \label{fig:phase_space_representative_phases}
  \end{figure*}

Fig.~\ref{fig:phase_space_representative_phases} presents representative snapshots of the main pattern families. The black arrows in each panel indicate the aggregate direction of motion, making it possible to distinguish flowing, stationary, and axis-dependent propagation states that may appear similar in a static spatial snapshot alone. Panels range from elongated polarized flows (a, b, f--h) through round and uniform states (d, j--l) to stationary clustered configurations (c, g), illustrating the geometric diversity captured by the classifier across the parameter sweep.

  \subsection*{Collective Order Is Reflected Differently Across Controller Stages}

With the regimes identified, we next examine how collective organization relates to controller structure at local and global scales. At the interaction scale, local directional order is captured by neighborhood nematicity \cite{deGennes1993}. Let $\theta_{ij}^{(t)}$ denote the heading of the $j$-th neighbor of agent $i$ at time $t$. The local nematic order parameter is
{
\begin{equation}
S_{\mathrm{nem},i}^{(t)} =
\sqrt{
\left(
\frac{1}{\kappa}\sum_{j=1}^{\kappa}\cos\left(2\theta_{ij}^{(t)}\right)
\right)^2
+
\left(
\frac{1}{\kappa}\sum_{j=1}^{\kappa}\sin\left(2\theta_{ij}^{(t)}\right)
\right)^2
},
\label{eq:local_nematicity}
\end{equation}
}
averaged across agents and over the final $T{=}25$ saved frames to yield $\bar{S}_{\mathrm{nem}}^{\mathrm{local}}$. Nematicity is preferred over polarization because several recurrent regimes exhibit counter-flows or head-to-tail symmetry, under which polarization becomes artificially small.

At the system scale, global structural coherence is quantified through the spectral gap of a Gaussian distance-weighted interaction graph. A fully connected weighted adjacency matrix $W_{ij}^{(t)} = \exp(-\lVert \mathbf{x}_i^{(t)} - \mathbf{x}_j^{(t)} \rVert^2 / 2\sigma_t^2)$ is constructed from particle positions at each frame, with $\sigma_t$ set to the maximum nearest-neighbor distance, a choice ensuring that the weighted graph remains sensitive to all interacting clusters present in the configuration. The global order variable is the first nontrivial spectral gap of the corresponding normalized graph Laplacian \cite{Chung1997},
{
\begin{equation}
\psi_{\mathrm{global}}^{(t)} = \frac{\lambda_1^{(t)} - \lambda_0^{(t)}}{2},
\label{eq:psi_global}
\end{equation}
}
where $\lambda_0^{(t)}$ and $\lambda_1^{(t)}$ are its two smallest eigenvalues. Small values indicate bottlenecks or near-fragmentation; larger values indicate a globally coherent interaction structure. The mature-window average $\bar{\psi}_{\mathrm{global}}$ is used throughout. 
  
A third spatial descriptor is the L\'{o}pez-Ruiz-Mancini-Calbet (LMC) complexity \cite{LopezRuiz1995} of the angular heading distribution, denoted $C_{\mathrm{LMC}}^{\angle}$, which captures the structured variability of the local sensory environment. Let $p_m$ denote the empirical probability mass in bin $m$ after partitioning the sampled angular values into $M$ intervals. The normalized Shannon entropy and disequilibrium are
\begin{equation}
H = -\frac{1}{\log M}\sum_{m=1}^{M} p_m \log p_m, \qquad
D = \sum_{m=1}^{M}\left(p_m - \frac{1}{M}\right)^2,
\label{eq:lmc_entropy_diseq}
\end{equation}
and the corresponding LMC complexity is $C_{\mathrm{LMC}} = H \cdot D$. This quantity is small for both narrowly concentrated and nearly uniform distributions, and peaks for structured intermediate regimes. Applied to the local heading distribution, it provides a descriptor of the sensory-environment complexity encountered by the agents.

On the neural side, the controller is separated into two functional stages. For the output stage, a linear surrogate $\hat{Y} = f_{\vartheta}(\mathbf{v})$ is fitted to the learned input--output map and evaluated using the coefficient of determination $R^{2}$. The proxy $1-R^{2}$ measures output nonlinearity: values near zero indicate an approximately linear steering policy, while larger values indicate nonlinear departure. More complex information-theoretic descriptors \cite{Tishby1999} could capture richer aspects of the input--output relationship; the linear proxy was chosen for interpretability. For the hidden layer, we apply the same LMC construction to the empirical distribution of sampled activation responses, yielding the hidden-layer complexity $C_{\mathrm{LMC}}$.

\begin{figure*}[!h]
  \centering
    \begin{minipage}[t]{0.49\textwidth}
        \centering
        \includegraphics[width=\linewidth]{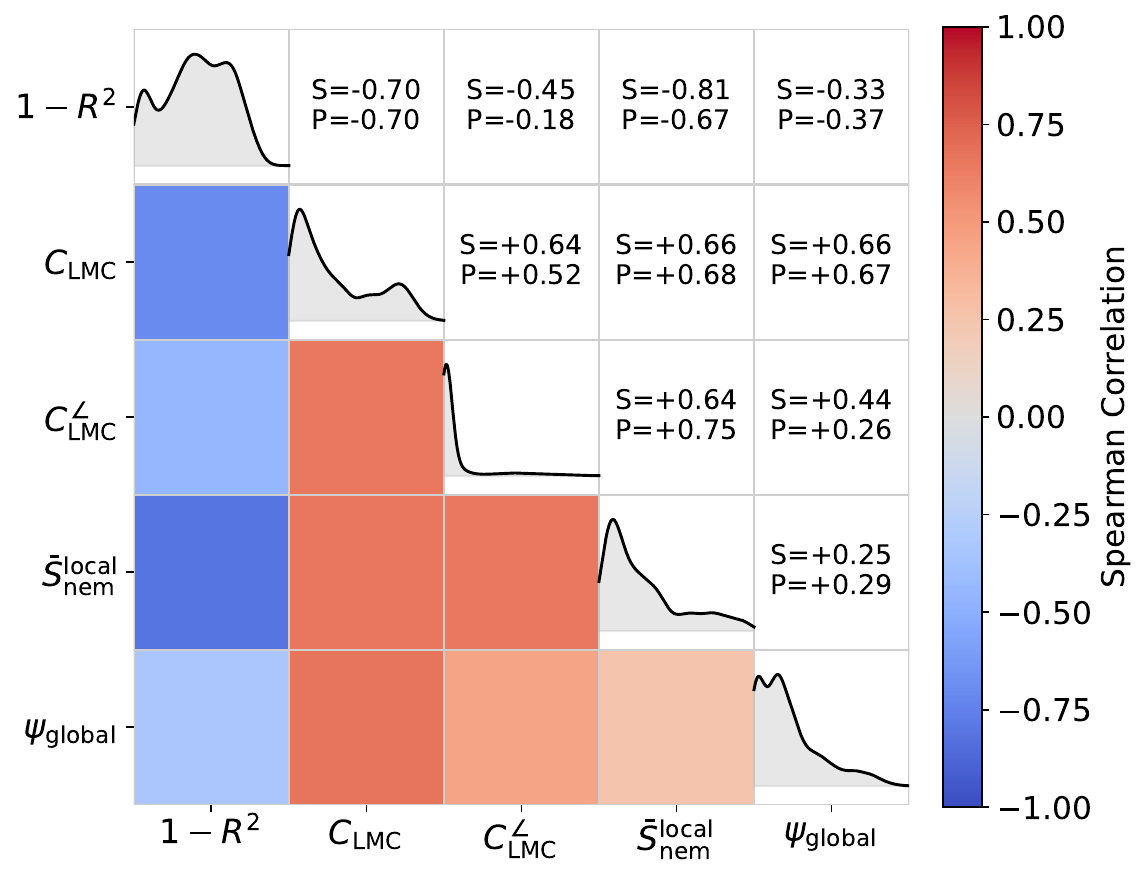}
        \par\vspace{2pt}
        \small \textbf{A} Spearman and Pearson association matrix for the five principal measures: neural nonlinearity ($1-R^2$), hidden-layer complexity ($C_{\mathrm{LMC}}$), angular complexity ($C^{\angle}{\mathrm{LMC}}$), local nematic order ($\bar{S}^{\mathrm{local}}{\mathrm{nem}}$), and global order ($\psi_{\mathrm{global}}$). Diagonal panels show kernel density estimates of the marginal distribution of each measure. Lower triangle: color-coded by Spearman correlation. Upper triangle: Spearman (S) and Pearson (P) coefficients.
    \end{minipage}
    \hfill
    \begin{minipage}[t]{0.49\textwidth}
        \centering
        \includegraphics[width=\linewidth]{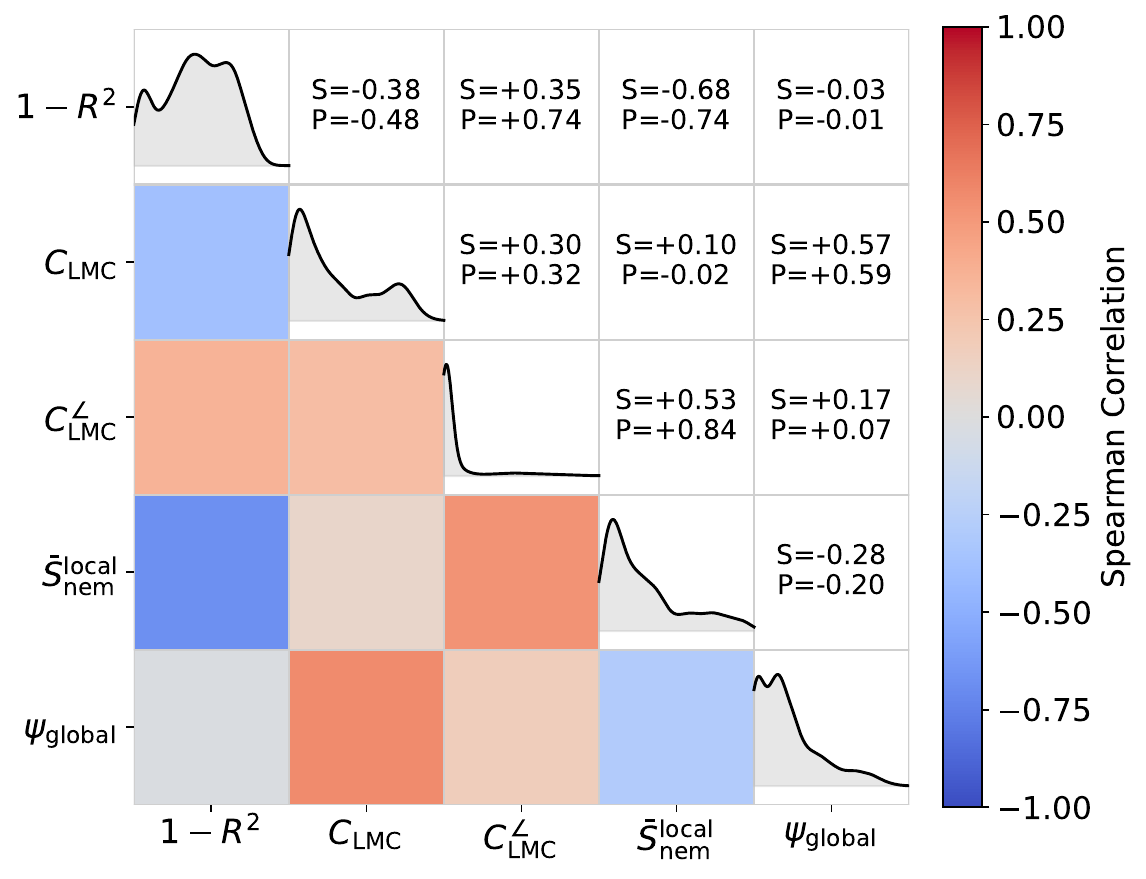}
        \par\vspace{2pt}
        \small \textbf{B} Partial-association matrix for the same five measures, controlling each pair for the remaining three variables. Diagonal panels again show kernel density estimates of the marginal distribution of each measure for reference. Lower triangle: color-coded by partial Spearman correlation. Upper triangle: partial Spearman (S) and partial Pearson (P) coefficients.
    \end{minipage}
    \caption{Broad and residual association structure among the five principal neural and collective descriptors. Panel \textbf{A} shows the full pairwise correlation pattern. Panel \textbf{B} shows the residual structure after controlling for the remaining variables. Diagonal panels show kernel density estimates of each variable's marginal distribution.}
    \label{fig:corr_compare_wide}
\end{figure*}

Using these five descriptors, we compute pairwise Pearson \cite{Pearson1895} and Spearman \cite{Spearman1904} correlations across all non-disordered runs. The resulting association matrix (Fig.~\ref{fig:corr_compare_wide}A) shows a clear separation in sign structure. Hidden-layer complexity is positively associated with both spatial order measures and with angular complexity, whereas output nonlinearity is negatively associated with those same variables. The diagonal panels show kernel density estimates of each variable's marginal distribution. Panel B shows the partial-association structure after controlling each pair for the remaining three variables, previewing the scale-resolved analysis that follows.

Reading the coefficients from Fig.~\ref{fig:corr_compare_wide}A, the strongest monotonic association is between output nonlinearity and local order ($\rho_S = -0.806$): as local organization increases, the controller output becomes progressively more linear. Output nonlinearity is also negatively correlated with hidden-layer complexity ($\rho_S = -0.696$). In contrast, hidden-layer complexity is positively correlated with local order ($\rho_S = 0.655$), global order ($\rho_S = 0.658$), and angular complexity. Together, these associations indicate that more ordered configurations combine more linear output mappings with greater hidden-layer complexity, and that this hidden-layer complexity tracks the structured variability of the sensory environment encountered by the agents. This latter relation is consistent with the law of requisite complexity \cite{Boisot2011}, a reformulation of Ashby's law of requisite variety \cite{Ashby1956}. Unevolved control simulations reported in SI Appendix further sharpen this interpretation: the hidden-layer and angular-complexity relations collapse without evolution, indicating that this representational-complexity pattern is produced by the evolutionary process.

The unevolved controls also clarify the interpretation of the output-stage result. The strong association between output linearity and local order (Fig.~\ref{fig:corr_compare_wide}A) persists even without evolutionary adaptation, showing that the constrained environment is itself structured to favor linear input--output mappings for producing local order. This corresponds to the ecological component of Gigerenzer's ecological rationality framework \cite{Gigerenzer1999}: the environment's informational structure makes linear strategies effective independently of whether agents are adapted toward them. In the evolved system, adaptation then amplifies this environmental regularity by preferentially retaining more linear controllers, adding the rationality component to the ecological precondition. A supplementary experiment using two-hidden-layer controllers confirms that this output-linearity association is not specific to single-hidden-layer topology (SI Appendix).

\subsection*{The Two Controller Stages Respond to Different Spatial Scales}

The broad correlations in Fig.~\ref{fig:corr_compare_wide}A do not show whether the observed effects share the same phenomenological origin or instead arise from distinct sources. Fig.~\ref{fig:corr_compare_wide}B addresses this by showing the partial-association matrix after controlling each pair for the other three descriptors. This distinction is important because persistent coupling would suggest that the reported trends largely reflect a common origin, potentially including architectural or topological effects within the controller, whereas selective uncoupling would support a partitioned organization in which different controller components coevolve while contributing differently to collective behavior.

Table~\ref{tab:partial_correlations} quantifies the intermediate step of this separation by controlling for a single spatial variable at a time (Materials and Methods).

\begin{table}[!h]
    \centering
    \caption{Partial Spearman correlations between the main neural and spatial descriptors, controlling for the complementary spatial variable.}
    \label{tab:partial_correlations}
    \footnotesize
    \setlength{\tabcolsep}{5pt}
    \begin{tabular}{lcc}
        \hline
        Association & Control & Partial Spearman \\
        \hline
        $1-R^2$ vs. $\bar{S}_{\mathrm{nem}}^{\mathrm{local}}$ & $\psi_{\mathrm{global}}$ & $-0.792$ \\
        $1-R^2$ vs. $\psi_{\mathrm{global}}$ & $\bar{S}_{\mathrm{nem}}^{\mathrm{local}}$ & $-0.228$ \\
        $C_{\mathrm{LMC}}$ vs. $\bar{S}_{\mathrm{nem}}^{\mathrm{local}}$ & $\psi_{\mathrm{global}}$ & $0.675$ \\
        $C_{\mathrm{LMC}}$ vs. $\psi_{\mathrm{global}}$ & $\bar{S}_{\mathrm{nem}}^{\mathrm{local}}$ & $0.677$ \\
        $C_{\mathrm{LMC}}$ vs. $1-R^2$ & $\bar{S}_{\mathrm{nem}}^{\mathrm{local}}$ & $-0.376$ \\
        $C_{\mathrm{LMC}}$ vs. $1-R^2$ & $\psi_{\mathrm{global}}$ & $-0.674$ \\
        \hline
    \end{tabular}
\end{table}

Comparing the first two rows of Table~\ref{tab:partial_correlations}, output nonlinearity remains strongly tied to local order after controlling for global order ($-0.792$), but its association with global order weakens substantially after controlling for local order ($-0.228$). The third and fourth rows show that hidden-layer complexity, by contrast, remains equally associated with both spatial scales after controlling for the other ($0.675$ and $0.677$). The final two rows confirm that the two neural descriptors remain negatively associated after controlling for either spatial scale, so the division between richer hidden representations and simpler output mappings is not reducible to a single shared spatial correlate.

When all remaining variables are controlled simultaneously, as shown in Fig.~\ref{fig:corr_compare_wide}B, the scale separation sharpens into an almost complete dissociation. Reading the partial coefficients from Panel B: output nonlinearity becomes exclusively local (partial $\rho_S = -0.68$ with $\bar{S}_{\mathrm{nem}}^{\mathrm{local}}$, $-0.03$ with $\psi_{\mathrm{global}}$), while hidden-layer complexity becomes exclusively global (partial $\rho_S = 0.57$ with $\psi_{\mathrm{global}}$, $0.1$ with $\bar{S}_{\mathrm{nem}}^{\mathrm{local}}$). The output stage is therefore shaped primarily by the interaction-scale organization directly relevant to steering, while the internal representation tracks the larger-scale organization of the collective. 

Bootstrap resampling ($10^4$ resamples with replacement) confirms that all main Spearman correlations are stable, with narrow $95\%$ confidence intervals excluding zero: $[-0.821, -0.789]$ for $1-R^2$ versus $\bar{S}_{\mathrm{nem}}^{\mathrm{local}}$, $[0.631, 0.678]$ for $C_{\mathrm{LMC}}$ versus $\bar{S}_{\mathrm{nem}}^{\mathrm{local}}$, and $[0.632, 0.682]$ for $C_{\mathrm{LMC}}$ versus $\psi_{\mathrm{global}}$.

\subsection*{The Pattern Is Preserved at the Regime Level}

\begin{figure*}[!h]
\centering
    \includegraphics[width=0.83\linewidth, height=0.8\textwidth]{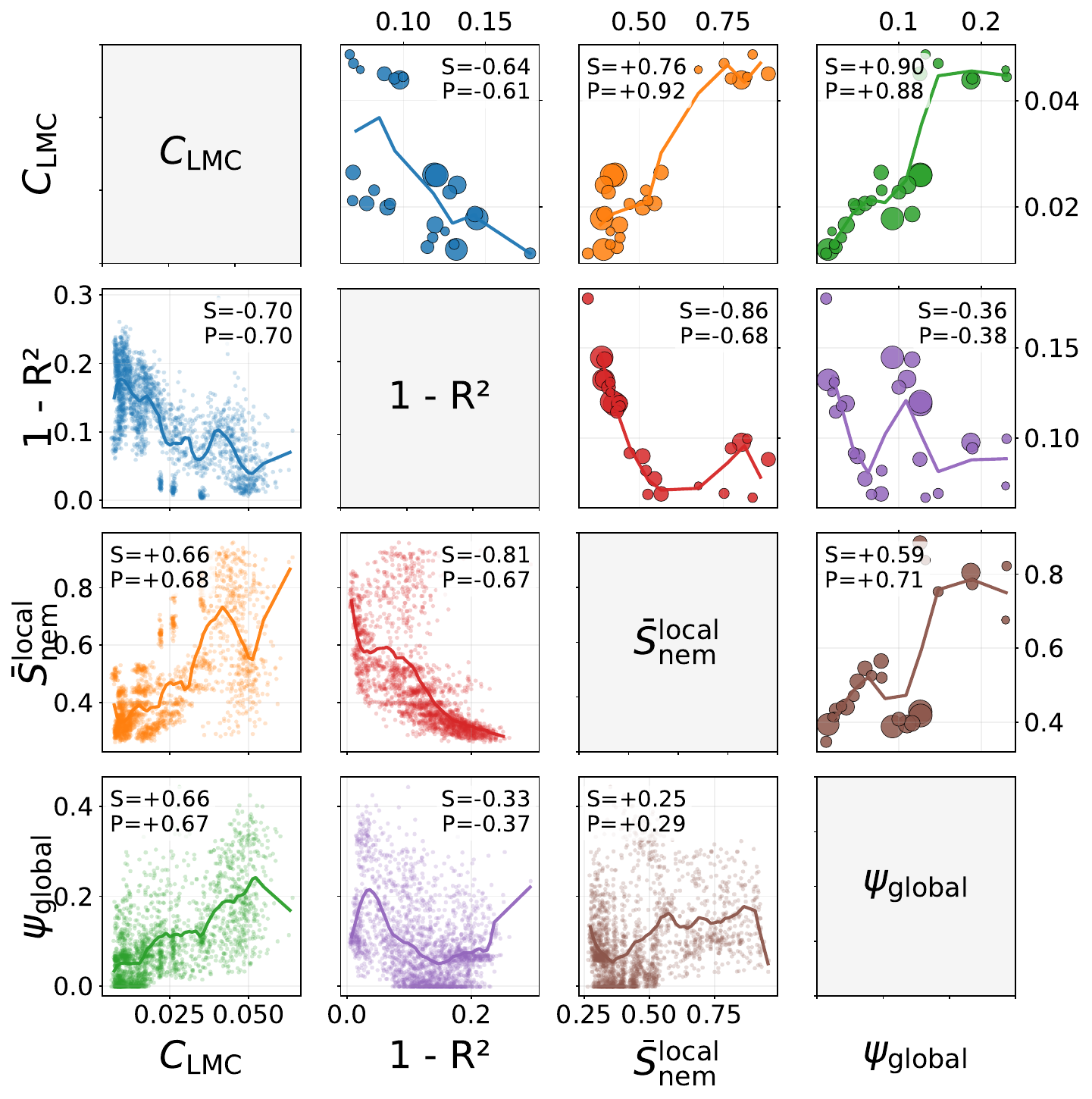}
    \caption{Combined pairwise comparison matrix for hidden-layer complexity ($C_{\mathrm{LMC}}$), neural nonlinearity ($1-R^2$), local nematic order ($\bar{S}_{\mathrm{nem}}^{\mathrm{local}}$), and global order ($\psi_{\mathrm{global}}$). Lower triangle: run-level scatter plots. Upper triangle: class-averaged pattern summaries. Each panel includes a smoothed trend line and Spearman (S) and Pearson (P) coefficients. Mirrored panels use the same color to indicate the same metric pair across the two aggregation levels.}
    \label{fig:cross_scattered_pairs}
\end{figure*}

To determine whether these associations are expressed across recurrent behavioral regimes, we aggregate the descriptors by classifier-defined pattern class and recompute the correlations using class-level means (Fig.~\ref{fig:cross_scattered_pairs}).

The visual comparison in Fig.~\ref{fig:cross_scattered_pairs} shows that the broad scatter visible in the lower-triangle panels tightens considerably in the upper-triangle panels, while the trend directions are preserved. The corresponding numerical values are reported in Table~\ref{tab:class_level_correlations}.

\begin{table}[!h]
    \centering
    \caption{Class-aggregated Pearson and Spearman correlations between the main neural and spatial descriptors.}
    \label{tab:class_level_correlations}
    \footnotesize
    \setlength{\tabcolsep}{5pt}
    \begin{tabular}{lcc}
        \hline
        Association & Pearson & Spearman \\
        \hline
        $1-R^2$ vs. $\bar{S}_{\mathrm{nem}}^{\mathrm{local}}$   & $-0.684$ & $-0.860$ \\
        $1-R^2$ vs. $\psi_{\mathrm{global}}$  & $-0.383$ & $-0.364$ \\
        $C_{\mathrm{LMC}}$ vs. $\bar{S}_{\mathrm{nem}}^{\mathrm{local}}$  & $0.923$ & $0.761$ \\
        $C_{\mathrm{LMC}}$ vs. $\psi_{\mathrm{global}}$ & $0.876$ & $0.901$ \\
        $C_{\mathrm{LMC}}$ vs. $1-R^2$        & $-0.610$ & $-0.636$ \\
        \hline
    \end{tabular}
\end{table}

Several associations strengthen substantially after aggregation (Table~\ref{tab:class_level_correlations}), with the Spearman correlation between $C_{\mathrm{LMC}}$ and $\psi_{\mathrm{global}}$ rising from $0.658$ to $0.901$ and $C_{\mathrm{LMC}}$ versus $\bar{S}_{\mathrm{nem}}^{\mathrm{local}}$ reaching a Pearson value of $0.923$, indicating that the signal is carried by qualitative differences between recurrent regimes rather than introduced by aggregation. In that setting, the controllers that evolve the most linear output mappings are the same ones that produce the most locally ordered collective configurations, consistent with Gigerenzer's ecological rationality framework \cite{Gigerenzer1999}.

\section*{Discussion}
\label{sec:Discussion}

Because fitness is defined from a physical property of the collective state, neural adaptation and collective organization are reflexively coupled in the present system. The analysis therefore does not establish causal directionality between controller structure and collective order. What it does establish is a two-part statistical pattern: ordered regimes combine richer hidden representations with more linear output mappings, and hidden-layer complexity covaries with the angular complexity of the local sensory environment. The first component is consistent with the law of requisite complexity \cite{Boisot2011}, and is not an artifact of the shallow architecture: unevolved controls show the representational-complexity relations collapse without optimization, and a separate study using LLM-generated coordination code \cite{cunha2026llm} reproduces the same qualitative relationship across a fundamentally different representational substrate. The second component, the progressive linearization of the output mapping with increasing collective order, points in a different theoretical direction. The present system operates under limited sensory access, stochastic perturbation, and bounded actuation, the conditions that define the Simon–Gigerenzer setting \cite{Simon1956,Gigerenzer1999}. The unevolved controls reveal that this linearization pattern has two layers: the constrained environment is itself structured such that more linear controllers produce more locally ordered configurations, even without adaptation. This is the ecological precondition of Gigerenzer's framework. Evolution then exploits this environmental regularity by selecting toward linearity, adding the adaptive component. The two layers together constitute a concrete instance of ecological rationality decomposed into its environmental and adaptive parts.

Related lines of work have connected adaptive and intelligent systems either to Ashby-style requirements on internal complexity \cite{Ashby1956,Friston2024,Seth2002} or to Gigerenzer-style ecological rationality, in which simpler decision rules can remain effective under constraint \cite{Chater2003,GigerenzerBrighton2009,Todd2007}. To our knowledge, these frameworks have rarely been brought together within the same system and interpreted as acting at different computational stages, which is the central empirical contribution of the present work. The evidentiary basis for this claim follows the same approach used to establish the original frameworks: Boisot and McKelvey supported requisite complexity through organizational case studies and conceptual argument \cite{Boisot2011}, and Gigerenzer established ecological rationality through empirical comparisons of heuristic performance on prediction tasks \cite{Gigerenzer1999}. What remains open is whether the observed structure depends on controller architecture. The scale separation could reflect properties of shallow networks processing local inputs, and whether the same two-stage organization appears in deeper controllers has yet to be tested.

Taken together, the results suggest that these two apparently opposing principles coexist by acting at different controller stages. Requisite complexity appears at the representational stage, whereas ecological rationality appears at the output stage. The full-control partial correlations further show that this division aligns with spatial scale, with the hidden layer coupled to global order and the output layer coupled to local order. In this view, adaptive controllers do not need to become uniformly more complex. Instead, complexity can concentrate where the environment must be represented, while action selection becomes comparatively simple. A similar division is familiar in cognitive systems: humans and other animals possess highly complex internal machinery, yet often simplify decisions to near-linear or binary choices \cite{Iyengar2000}. What we observe here may likewise reflect a form of computational efficiency, which would make sense in systems operating under informational or energetic constraint. That interpretation is consistent with Simon's bounded-rationality framework, which preceded Gigerenzer's ecological rationality account and emphasized decision-making under limited information and finite resources \cite{Simon1956,Gigerenzer1999}. Although the present analysis does not establish a common mechanism, the parallel suggests that separating internal representational complexity from action-level simplification may be a recurring strategy in adaptive systems.

\section*{Materials and Methods}

\subsection*{Parameter Sweep and Training Horizon}
The simulations sweep over four primary sensory and actuation constraints: field of view (FoV), output noise $\xi$, maximum turning angle $\omega_{\text{max}}$, and topological neighborhood size $\kappa$ (Table~\ref{tab:sec3-parameters_table}). The remaining model parameters, including the evolutionary algorithm and its hyperparameters, are held fixed; a full description is provided in SI Appendix. The full sweep comprises 3024 simulations spanning the Cartesian product of the swept ranges.

\begin{table}[!h]
      \caption{Parameters used in the swarm analyses.}
      \label{tab:sec3-parameters_table}
      \centering
      \footnotesize
      \setlength{\tabcolsep}{3pt}
      \begin{tabular}{| c | c | c |}
          \multicolumn{3}{l}{\textit{Fixed parameters}} \\
          \hline
          Parameter & Name & Value \\
          \hline
          $N$      & Number of agents   & $100$ \\
          $L^2$    & Lattice area       & $20 \times 20$ area units \\
          $\rho$   & Agent density      & $0.25$ \\
          $v_{0}$  & Agent speed        & $1$ length units $(\text{time units})^{-1}$ \\
          $\Delta t$ & Time step         & $0.1$ \\
          $R$      & Collision radius   & $0.2$ length units \\
                   & Boundary           & Toroidal (periodic) \\
          \hline
          \noalign{\vskip 5pt}
          \multicolumn{3}{l}{\textit{Evolutionary parameters}} \\
          \hline
          Parameter & Name & Value \\
          \hline
          $T_{\text{repr}}$ & Reproduction interval & $300$ steps \\
          $\sigma_{\text{share}}$ & Sharing threshold & $0.67\,\langle \Delta_{ij} \rangle$ \\
          $\alpha$  & Sharing exponent    & $1$ \\
          $\mu$     & Mutation rate        & $0.1$ \\
          $\eta$    & Mutation strength    & $0.1$ \\
                    & Selection            & Roulette (fitness-proportionate) \\
                    & Recombination        & Uniform crossover \\
          \hline
          \noalign{\vskip 5pt}
          \multicolumn{3}{l}{\textit{Swept parameters (3024 runs)}} \\
          \hline
          Parameter & Name & Values \\
          \hline
          FoV               & Field of view       & $\pi/2,\;\pi,\;3\pi/2,\;2\pi$ \\
          $\omega_{\text{max}}$ & Max.\ turning angle & $\pi/2,\;\pi,\;3\pi/2,\;2\pi$ \\
          $\kappa$          & Nearest neighbors   & $1,2,\dots,9$ \\
          $\xi$             & Output noise        & 21 levels in $[0,\,1]$\textsuperscript{$\dagger$} \\
          \hline
      \end{tabular}

      \vspace{2pt}
      {\scriptsize \textsuperscript{$\dagger$}Non-uniformly spaced with finer sampling at low amplitudes; full list in SI Appendix.}
\end{table}

\begin{figure}[!ht]
    \centering
    \includegraphics[width=0.9\linewidth]{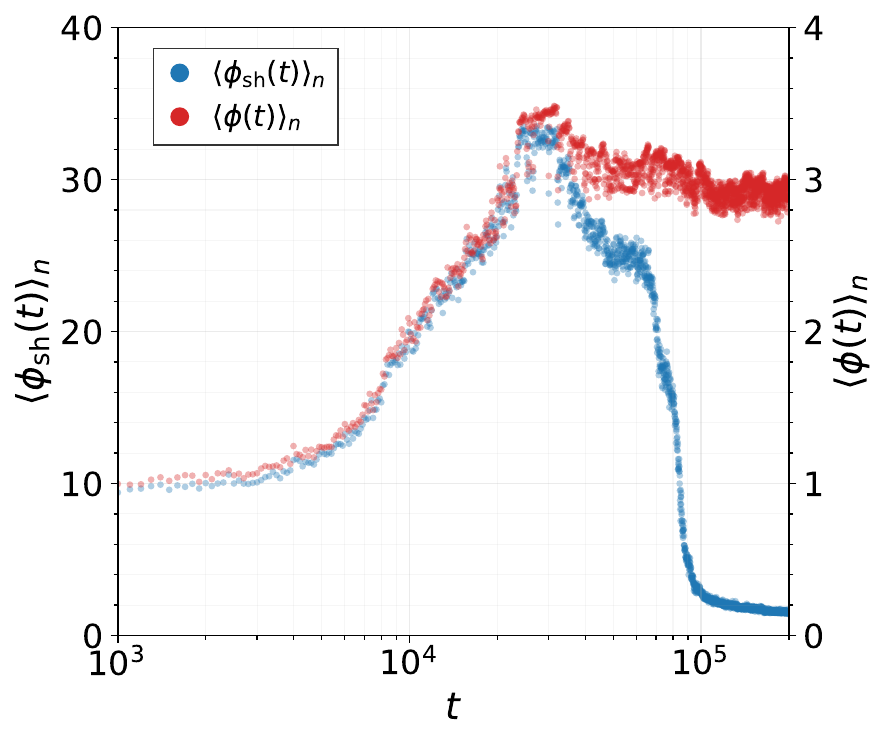}
    \caption{Average fitness ($\langle\varphi(t)\rangle$) and average shared fitness ($\langle\varphi_{sh}(t)\rangle$) as a function of training steps.}
    \label{fig:sec3-Fitness_Evolution}
\end{figure}

Fitness trajectories stabilize before $1.5\times 10^{5}$ simulation steps across all parameter ranges (Fig.~\ref{fig:sec3-Fitness_Evolution}), confirmed by tracking both the average fitness $\langle \varphi(t)\rangle$ and the diversity-corrected shared fitness $\langle \varphi_{sh}(t)\rangle$ \cite{FLOREANO2023}. We adopt this horizon as a conservative training cutoff and extract all subsequent measurements from the final 25 saved frames.

\subsection*{Geometric Classifier}
The classifier assigns compositional labels to each simulation through a hierarchical sequence of geometric tests. First, a Gaussian-blurred occupancy field determines whether the agent distribution is spatially homogeneous or concentrated into localized density peaks. When no peaks are detected, the label is resolved from orientational order alone. When peaks are present, the classifier determines whether they form a single connected basin or multiple disconnected basins, then refines each case through orientational order, center-of-mass motion (stationary vs.\ flowing), and shape anisotropy (round vs.\ elongated). For elongated configurations, the dominant propagation axis is also identified. The final label is composed of tokens produced at each stage (basin count, orientational order, motion state, shape, propagation direction), so that configurations differing in only one component are treated as neighbors in the phase-space analysis. Labels are assigned by majority vote over the final frames to suppress transient fluctuations. Full classifier details are provided in SI Appendix.

\subsection*{Phase-Space Interpolation}
The phase-space field shown in Fig.~\ref{fig:phase_space_classification} is obtained by semantic composition interpolation between sampled classifier labels, preserving similarities among regimes that differ by only a subset of label components; full interpolation details are provided in SI Appendix.

\subsection*{Descriptor Estimation and Analysis Sample}

\paragraph*{LMC complexity estimates}
Angular behavioral complexity $C_{\mathrm{LMC}}^{\angle}$ is estimated from an adaptive one-dimensional histogram of the local heading distribution, using $\min(32,\mathrm{round}(\sqrt{N}))$ bins, where $N$ is the number of agents. Hidden-layer complexity $C_{\mathrm{LMC}}$ is estimated from an adaptive three-dimensional histogram of the sampled hidden activations after principal-component projection, using $\min(16,\mathrm{round}(n_{\mathrm{samples}}^{1/5}))$ bins per axis. Full implementation details, including the PCA projection and histogram construction, are provided in SI Appendix.

\paragraph*{Output nonlinearity}
Output nonlinearity is quantified by fitting a linear surrogate of the form $Y = A X + B$ to the controller input--output map and evaluating the coefficient of determination $R^2$. The proxy $1-R^2$ is then used as the measure of departure from linear structure.

\paragraph*{Bootstrap analysis}
Bootstrap confidence intervals were computed from $10^4$ resamples with replacement \cite{Efron1979}; the full bootstrap results are reported in SI Appendix.

\paragraph*{Analysis sample sizes}
The full parameter sweep contains 3024 runs. Correlations involving the linear surrogate exclude the 762 runs classified as disordered, leaving 2262 runs in the main run-level statistical analysis. At the class-aggregated level, the geometric classifier produces 26 recurrent classes, so each class-level statistic is computed from 26 points. Class sizes are heterogeneous; the full class counts are reported in SI Appendix.

\section*{Acknowledgment}
\acknow{We would like to thank the School of Informatics, Computing, and Cyber Systems and the College of Engineering, Informatics, and Applied Sciences for supporting this work.}

\section*{Disclaimer}
The authors declare no conflict of interest. This work was not supported by any external funding.

\showacknow{} % Display the acknowledgments section

% \bibsplit[3]
%Use \bibsplit to split the references from the body of the text. Value "[3]" represents the number of reference in the left column (Note: Please avoid single column figures & tables on this page.)

\bibliography{collectivity}

\end{document}